\documentclass[conference, letterpaper]{IEEEtran}
\ifCLASSINFOpdf
\else
\fi
\usepackage{subcaption}

\ifCLASSINFOpdf
   \usepackage[pdftex]{graphicx}
\else
\fi

\usepackage[cmex10]{amsmath}
\usepackage{color}
\usepackage{fancyhdr}

\renewcommand{\thispagestyle}[2]{}

\fancypagestyle{plain}{
        \fancyhead{}
        \fancyhead[C]{first page center header}
        \fancyfoot{}
        \fancyfoot[C]{first page center footer}
}
\usepackage[ruled]{algorithm2e}
\usepackage{multirow}
\usepackage{amsmath}
\usepackage{pbox}
\usepackage{morefloats}
\usepackage{xcolor}
\usepackage{array}
\usepackage{amssymb}

\usepackage{fancyvrb}
\usepackage{threeparttable}
\usepackage{standalone}
\usepackage[justification=centering]{caption} 
\usepackage{url}

\begin{document}

%
\title{A Cost-Efficient Look-Up Table Based Binary Coded Decimal Adder Design}

\author{\IEEEauthorblockN{Zarrin~Tasnim~Sworna\IEEEauthorrefmark{1}, Mubin~Ul~Haque \IEEEauthorrefmark{1}, Hafiz Md. Hasan Babu\IEEEauthorrefmark{1} and Lafifa Jamal\IEEEauthorrefmark{1}}

\IEEEauthorblockA{\IEEEauthorrefmark{1}Department of Computer Science and Engineering, University of Dhaka, Dhaka-1000, Bangladesh.\\
Email:mubin10haque@gmail.com;sworna07@yahoo.com;hafizbabu@cse.univdhaka.edu;lafifa@yahoo.com\\ Corresponding author's email: hafizbabu@cse.univdhaka.edu, hafizbabu@du.ac.bd}
}


%


\maketitle

\begin{abstract}

The BCD (Binary Coded Decimal) being the more accurate and human-readable representation with ease of conversion, is prevailing in the computing and electronic communication. In this paper, a tree-structured parallel BCD addition algorithm is proposed with the reduced time complexity $O(N(\log_2b)+(N-1))$, where $N$ = number of digits and $b$ = number of bits in a digit. BCD adder is more effective with a LUT (Look-Up Table)-based design, due to FPGA (Field Programmable Gate Array) technology's enumerable benefits and applications. A size-minimal and depth-minimal LUT-based BCD adder circuit construction is the main contribution of this paper. The proposed parallel BCD adder gains a radical achievement compared to the existing best-known LUT-based BCD adders. The proposed BCD Adder provides prominent better performance with 20.0\% reduction in area and 41.32\% reduction in delay for the post-layout simulation. Since the proposed circuit is improved both in area and delay parameter, it is 53.06\% efficient in terms of area-delay product compared to the best known existing BCD adder, which is surely a significant achievement.

\end{abstract}



%
\IEEEpeerreviewmaketitle

\section{Introduction}
BCD (Binary Coded Decimal) representation is advantageous due to its finite place value representation, rounding, easy scaling by a factor of 10, simple alignment and conversion to character form \cite{1} \cite{22}. It is highly used in embedded applications,  digital communication and financial calculations \cite{23} \cite{24}. Hence, faster and efficient BCD addition method is desired. In this paper, a $N$-digit addition method is proposed which omits the complex manipulation steps, reducing area  and delay of the circuit. The application of FPGA in cryptography, NP (Non Polynomial)-Hard optimization problems, pattern matching, bioinformatics, floating point arithmetic, molecular dynamics is increasing radically \cite{21} \cite{25} \cite{26}. Due to re-configurable capabilities, FPGA implementation of BCD addition is of concern. LUT being one of the main components of FPGA, a LUT-based adder circuit is proposed.\\\\
Two main contributions are addressed in this paper. Firstly, a new tree-based parallel BCD addition algorithm is presented. Secondly, a compact and high-speed BCD adder circuit with an improvement in time complexity of $O(N(\log_2b)+(N-1))$ is proposed, where $N$ represents the number of digits and $b$ represents the number of bits in a digit.\\
 
The organization of this paper is as follows: In the next section, the earlier approaches and their limitations are described. In Section \ref{sec: proposed}, a novel BCD addition method  is proposed. Then, the construction of BCD adder circuit is given. In Section \ref{sec: simulation}, the simulation results and performance analysis of the proposed circuit are elucidated. Last of all, the paper is concluded in Section \ref{sec: concl}.

\section{Literature Overview}

In this section, various types of the latest existing LUT-based BCD adders are presented. 

\subsection{Existing LUT-Based BCD Adders}
BCD adder uses BCD numbers as input and output \cite{6}. Since a 4-bit binary code has 16 different binary combinations, the addition of two BCD digits may produce incorrect result that exceeds the largest BCD digit $(9)_{10} = (1001)_{BCD}$ \cite{6} [9] [10]. In such cases, the result must be corrected by adding $(6)_{10} = (0110)_{BCD}$ to guarantee that the result is a BCD digit. The resultant decimal carry output generated by the correction process is added to the next higher digit of the BCD addends.\\

Authors in paper [9] proposed a direct implementation of BCD adder circuit. They had proposed two different architectures for the construction of LUT-based BCD adder circuit. A truth table had been formed for each input/output combination and the corresponding circuit was proposed in first architecture. It consumed eleven 6-input LUTs. Two level of abstraction was performed for the second architecture. First least two significant input bits were fed into the first level of circuits whereas the rest input bits along with the output of the first level were provided to the second level of the circuits. The second approach required seven number of 6-input LUTs with a much delay. The direct implementation suffers a significant LUT-delay product.\\

The BCD adder proposed in [10] used Virtex-6 platform to implement their circuit architecture which had been proposed earlier in [11]. Gao et al. proposed a BCD adder, where the first bit of the addends are added using a full adder and the most significant three bits are added using 6-input LUTs [11]. A correction is ensured in 6-input LUTs by adding $(3)_{10}$ to the sum if the sum of the most significant three bits is greater than or equals to $(5)_{10}$. Moreover, extra circuits are required, when the sum of the most significant three bits is $(4)_{10}$ and the carry generated from the full adder is one. The circuit being serial in architecture except the LUTs portions, requires much time complexity and delay which hinders faster output generation [11]. Bioul et al. proposed a BCD adder, where additions were performed in a carry chain type fashion and thus, suffered from a significant amount of delay [12]. They used Virtex-4 and Virtex-5 platform to show that the area overhead (in terms of required number of LUTs) with respect to binary computation is not negligible and it is around five times in Virtex-4 and nearly four times in Virtex-5. The main reason of such difference is due to the more complex definition of the carry propagate and carry generate functions. 

Authors in [13] proposed a BCD addition method, where six is added as a correction factor, when the sum of $A_i^U + B_i^U$ equals or greater than 8, where $A_i^U$ and $B_i^U$ represent the most significant three bits of the input operands $A$ and $B$, respectively. If the final output is (111)$_2$, then a replacement of (111)$_2$ with (100)$_2$ is required as a final step for the exact BCD output. Vazquez et al. presented various carry chain BCD addition methods and their implementations on  the LUT architecture [14]. As the carry-chain mechanisms being serial in architecture, the proposed methods in [14] require much delay which are surely a huge drawback. 

A power and area-efficient BCD adder circuit was proposed by the authors in paper [15]. They actually used the circuit architecture exhibited in [1] and estimate the power consumption of the circuit. The delay has been calculated on a Virtex-5 platform by using 6-input LUT and the value obtained was 6.22 ns. The average power consumption of the circuit described in [15] was 25 mW which achieved a significant improvement over conventional LUT-based BCD adder. However, the method proposed in [15] required a total of 48 logic elements which can be optimized further.

%
%

\label{sec: prior}

\section{Proposed Design of LUT-Based BCD Adder}
\label{sec: proposed}
In this section, firstly a BCD addition algorithm is proposed. Then a new LUT-based BCD adder is constructed. Essential figures and lemmas are presented to clarify the proposed ideas.
\subsection{Proposed Parallel BCD Addition Method}
The carry propagation is the main cause of delay of BCD adder circuit, which gives BCD adder a serial architecture. As the reduction of delay is one of the most important factor for the efficiency of the circuit, carry propagation mechanism needs to be removed for faster BCD addition. In this paper, a highly parallel BCD addition method is proposed with a tree-structured representation with significant reduction of delay. The proposed BCD addition method has mainly two steps which are as follows:
\begin{itemize}
\item \emph{Bit-wise addition of the BCD addends produce the corresponding sum and carry in parallel. For the addition of first bit, the carry from the previous digit will be added too and the produced sum will be the direct first bit of the output.}
\item \emph{If the most significant carry bit is zero then, except the first sum and last carry bit, add the other sum and carry bits in pair in parallel; and if the sum is greater than or equals to five, add three to the result to obtain the correct BCD output.} 
\item \emph{If the most significant carry bit is one then, update the final output values according to Equation 1 and  2.}
\end{itemize}
Suppose, $A$ and $B$ be the two addends of a 1-digit BCD adder, where BCD representations of $A$ and $B$ are $A_4 A_3 A_2 A_1$ and $B_4 B_3 B_2 B_1$, respectively. The output of the adder will be a 5-bit binary number $\{ C_{out} S_3 S_2 S_1 S_0 \}$ , where $C_{out}$ represents the position of tens digit and $\{S_3 S_2 S_1 S_0\}$ symbolizes unit digit of BCD sum. $A_0$ and $B_0$ are added along with $C_{in}$ which is the carry from the previous digit addition. If it is the first digit addition, the carry will be considered as zero. The produced sum bit will be the direct first bit of the output. Other pairwise bits $(B_1,A_1)$, $(B_2,A_2)$, $(B_3,A_3)$ will be added simultaneously. The resultant sum and carry bits $(S^{\alpha}_3, C_2, S^{\alpha}_2, C_1, S^{\alpha}_1$ and $C_0)$ are added pairwise providing output \{$C^{\gamma}_{out}$ $S^{\gamma}_3$ $S^{\gamma}_2$ $S^{\gamma}_1$ \} and corrected by addition of three according to the following Equation \ref{eq: three} and Equation \ref{equ: four}:\\

\begin{center}
$C^{\gamma}_{out} S^{\gamma}_3 S^{\gamma}_2 S^{\gamma}_1$
\end{center}
\vspace{-2ex}
\begin{equation}
    =
    \begin{cases}
      (C^{\gamma}_{out} S^{\gamma}_3 S^{\gamma}_2 S^{\gamma}_1), & \text{if}\ C_3=0 \hspace{2pt} and \hspace{1pt} C^{\gamma}_{out} S^{\gamma}_3 S^{\gamma}_2 S^{\gamma}_1 < 5 \\
      (C^{\gamma}_{out} S^{\gamma}_3 S^{\gamma}_2 S^{\gamma}_1)+3, & \text{if}\ C_3=0 \hspace{1pt} and \hspace{1pt} C^{\gamma}_{out} S^{\gamma}_3 S^{\gamma}_2 S^{\gamma}_1 \geq 5 \\
      1C_0S^{\beta}_2S^{\beta}_1, & \text{otherwise}\\
    \end{cases}    
\label{eq: three} 
  \end{equation}
  
\begin{equation}
\label{equ: four}
\text{where}\hspace{15pt}
      S^{\beta}_1=S^{\beta}_2=
      \begin{cases}
      0, & \text{if}\ C_0=1 \\
      1, & \text{otherwise}
      \end{cases}
\end{equation}

In Table \ref{tab:truthac1}, the truth table is designed with $(A_3, A_2, A_1)$ and $(B_3, B_2, B_3)$ as input and $(C_{out}$ $S_3$ $S_2$ $S_1)$ as the final BCD output by following required correction. $(S^{\alpha}_3$, $C_2$, $S^{\alpha}_2$, $C_1, S^{\alpha}_1$ and $C_0)$ are added pairwise as intermediate step, producing $(F_4, F_3, F_2, F_1)$ by considering carry $C_0$ always 1. A numeric $3 ((011)_2)$ is added to the intermediary output $F$, if $F$ is greater than or equals to five. A similar table considering $C_0$ as 0 can be calculated which is shown in Table \ref{tab:truthac0}. The truth tables verify the functions of each output of the LUTs of the BCD adder. The algorithm of $N$-digit BCD addition method is presented in Algorithm \ref{alg:Addition}.


\begin{table*}%

\centering

\caption{The Truth Table of 1-Digit BCD Addition with $C_0=1$\label{tab:truthac1}}{%

\scalebox{.9}{

\begin{tabular}{|l|l|l|l|l|l|l|l|l|l|l|}

\hline
$B(3:1)$     & ${A(3:1)}$ & $S^{\alpha}(3:1)$ & ${C(3:1)}$ & $C_0$    & $F(4:1)$ & Add 3& $C_{out}$ & $S_3$ & $S_2$ & $S_1$\\\hline 
000	&001	&001&	000&1& 	0010    &-&	    0&0&1&0\\\hline
000	&010	&010&	000&1&	0011	&-&	    0&0&0&0\\\hline
000	&011	&011&	000&1&	0100	&-&	    0&0&0&0\\\hline
000	&100	&100&	000&1&	0101	&Add 3&	1&0&0&0\\\hline	
001	&001	&000&	001&1&	0011	&-&	    0&0&0&0\\\hline
001	&010	&011&	000&1&	0100	&-&	    0&0&0&0\\\hline
001	&011	&010&	001&1&	0101	&Add 3&	1&0&0&0\\\hline
001	&100	&101&	000&1&	0110	&Add 3&	1&0&0&1\\\hline
.	&.	&.&	.&	&&&.	&&	&\\\hline
.	&.	&.&	.&	&&&.	&&	&\\\hline
.	&.	&.&	.&	&&&.	&&	&\\\hline
100	&001	&101&	000&1&	0110	&Add 3&	1&0&0&1\\\hline
100	&010	&110&	000&1&	0111	&Add 3&	1&0&1&0\\\hline
100	&011	&111&	000&1&	1000	&Add 3&	1&0&1&1\\\hline
100	&100	&000&	100&1&	1001	&Add 3&	1&1&0&0\\\hline
																	
\end{tabular}}}
\begin{tablenotes}
    \item  \hspace{150 pt} `-' Represents ``No correction by adding 3 is required.''
    \end{tablenotes}
\end{table*}%


\begin{table*}%
\centering
\caption{The Truth Table of 1-Digit BCD Addition with $C_0=0$\label{tab:truthac0}}{
%
\scalebox{.9}{
\begin{tabular}{|l||l||l||l|l||l||l||l|l|l|l|}
\hline
$B(3:1)$     & ${A(3:1)}$ & {$S^{\alpha}(3:1)$} & ${C(3:1)}$ & $C_0$    & $F(4:1)$ & Add 3& $C_{out}$ & $S_3$ & $S_2$ & $S_1$\\\hline 
000	&001	&001&	000&0& 	0001    &-&	    0&0&0&1\\\hline
000	&010	&010&	000&0&	0010	&-&	    0&0&1&0\\\hline
000	&011	&011&	000&0&	0011	&-&	    0&0&1&1\\\hline
000	&100	&100&	000&0&	0100	&-&	    0&1&0&0\\\hline	
001	&001	&000&	001&0&	0010	&-&	    0&0&1&0\\\hline
001	&010	&011&	000&0&	0011	&-&	    0&0&1&1\\\hline
001	&011	&010&	001&0&	0100	&-&	    0&1&0&0\\\hline
001	&100	&101&	000&0&	0101	&Add 3&	1&0&0&0\\\hline
.	&.	&.&	.&	&&&.	&&	&\\\hline
.	&.	&.&	.&	&&&.	&&	&\\\hline
.	&.	&.&	.&	&&&.	&&	&\\\hline
100	&001	&101&	000&0&	0101	&Add 3&	1&0&0&0\\\hline
100	&010	&110&	000&0&	0110	&Add 3&	1&0&0&1\\\hline
100	&011	&111&	000&0&	0111	&Add 3&	1&0&1&0\\\hline
100	&100	&000&	100&0&	1000	&Add 3&	1&0&1&1\\\hline		
\end{tabular}}
}

\begin{tablenotes}
    \item  \hspace{150 pt} `-' Represents ``No correction by adding 3 is required.''
    \end{tablenotes}
\end{table*}%

\begin{algorithm}
\SetAlgoNoLine
\KwIn{Two $N$-digit BCD numbers $A = {A^N ... A^3 A^2 A^1}$ and $B = {B^N ... B^3 B^2 B^1}$ where $A^i = A^i_3 A^i_2 A^i_1 A^i_0$ and $B^i = B^i_3 B^i_2 B^i_1 B^i_0$ with $i = 1, 2, 3, ... , N$\;}
\KwOut{Sum, $S = {S^N ... S^3 S^2 S^1}$ where $S^i = { S^i_3 S^i_2 S^i_1 S^i_0 }$ with $i = 1, 2, 3, ... , N$  and Carry, $C$ = $C^N_{out}$\;}

	 $i \leftarrow 1$;\\
	 
\Repeat{$i = N$}{
     $S^i_0 \leftarrow A^i_0 \oplus B^i_0 \oplus C^i_{in}$ and $C^i_0 \leftarrow A^i_0.B^i_0.C^i_{in}$;\\
     $S^{i\alpha}_1 \leftarrow A^i_1 \oplus B^i_1 $ and $C^i_1 \leftarrow A^i_1.B^i_1$;\\
     $S^{i\alpha}_2 \leftarrow A^i_2 \oplus B^i_2 $ and $C^i_2 \leftarrow A^i_2.B^i_2$;\\
     $S^{i\alpha}_3 \leftarrow A^i_3 \oplus B^i_3 $ and $C^i_3 \leftarrow A^i_3.B^i_3$ in parallel;\\
     
	 \eIf {($C^i_3=1$)}{\text{ $S^{i\beta}_3 \leftarrow
	 C_0$; $C^{i\beta}_{out}\leftarrow 1$;}\\
	 \eIf{$C^i_0=0$}{\text{$S^{i\beta}_1\leftarrow 1$; $S^{i\beta}_2\leftarrow 1$;}}{\text{$S^{i\beta}_1 \leftarrow 0$;$S^{i\beta}_2\leftarrow 0$}}}
	 {
	   {$C^{i\gamma}_{out}$ $S^{i\gamma}_3$ $S^{i\gamma}_2$ $S^{i\gamma}_1$} $\leftarrow$ $(S^{i\alpha}_3 S^{i\alpha}_2 S^{i\alpha}_1) + (C^i_2 C^i_1 C^i_0)$;\\
	  \If{{$C^{i\gamma}_{out}$ $S^{i\gamma}_3$ $S^{i\gamma}_2$ $S^{i\gamma}_1$ }$ \geq 5$}
	  {\text{${C^{i\gamma}_{out} S^{i\gamma}_3 S^{i\gamma}_2 S^{i\gamma}_1}\leftarrow C^{i\gamma}_{out} S^{i\gamma}_3 S^{i\gamma}_2 S^{i\gamma}_1 + 3$;}
	 }
}
	\eIf{($C^i_3=1$)}{$S^i_1 \leftarrow S^{i\beta}_1; S^i_2 \leftarrow S^{i\beta}_2; S^i_3 \leftarrow S^{i\beta}_3; C^i_{out} \leftarrow C^{i\beta}_{out};$}
	{$S^i_1 \leftarrow S^{i\gamma}_1; S^i_2 \leftarrow S^{i\gamma}_2; S^i_3 \leftarrow S^{i\gamma}_3; C^i_{out} \leftarrow C^{i\gamma}_{out};$}
}

\caption{Proposed Algorithm for an $N$-digit Parallel BCD Addition}
\label{alg:Addition}
\end{algorithm}

\begin{figure*}
\centerline{\includegraphics[scale=.5]{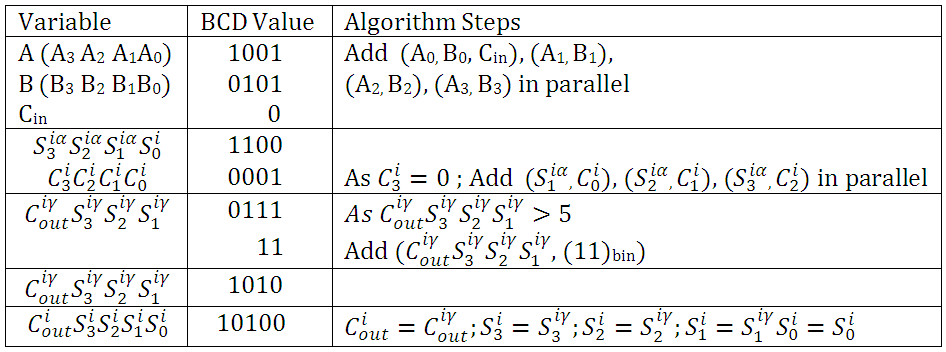}}
\caption{Example Demonstration of the Proposed BCD Addition Algorithm for $C^i_3=0$.}
\label{fig:ex1}
\end{figure*}

\begin{figure*}
\centerline{\includegraphics[scale=.5]{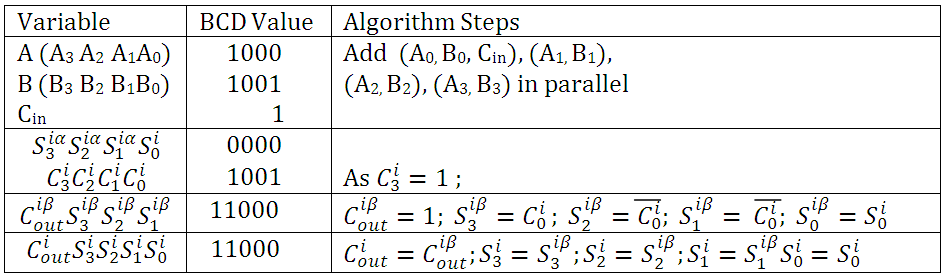}}
\caption{Example Demonstration of the Proposed BCD Addition Algorithm for $C^i_3=1$.}
\label{fig:ex2}
\end{figure*}

Two example of BCD addition method using the proposed algorithm is demonstrated in Fig. \ref{fig:ex1} and \ref{fig:ex2}, where $C^i_3=0$ and $C^i_3=1$, respectively. Each step of the example is mapped to the corresponding algorithm step for more clarification.  

\begin{figure}
\centerline{\includegraphics[scale=.5]{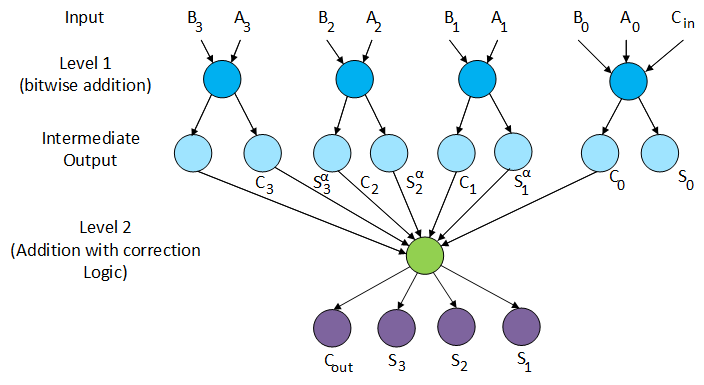}}
\caption{Tree Structure Representation of the Proposed BCD Addition Method.}
\label{fig:tree}
\end{figure}

The proposed BCD addition method can be represented as a tree-structure as it is parallel which is shown in Fig. \ref{fig:tree}. There are basically two operational levels of the tree. Starting from the inputs, in level 1, the bit-wise addition is performed and the intermediary resultants are obtained. Then, in level 2, the addition and correction are performed providing the correct BCD output. Hence, the time complexity of the proposed algorithm is logarithmic according to the operational depth of the tree. Lemma 3.1 is given to prove the time complexity of our proposed method. The time complexity of existing and proposed BCD adders are elucidated in Table \ref{tab:timc}.

\begin{table}%
\centering
\caption{Comparison of the Time Compleixities of the Proposed and Existing BCD Addition Methods\label{tab:timc}}{%
\begin{tabular}{|l||l|}
\hline
Method     & Time Complexity\\\hline 
Existing [11] & $O(N(b+2)+(N-1))$\\\hline
Existing [12] & $O(N(b+1)+(N-1))$\\\hline
Existing [13] & $O(N(b+3)+(N-1))$ \\\hline
Existing [14] & $O(N(b+3)+(N-1))$	\\\hline
Proposed &	$O(N(\log_2b)+(N-1))$ \\\hline
\end{tabular}}
\begin{tablenotes}%
\item `$b$':``number of bits in a digit'' and `$N$': ``number of digits''.
\end{tablenotes}%
\end{table}%

\emph{ \textbf{Lemma 3.1}
The proposed BCD addition algorithm requires at least $O(N(log_2b) + (N - 1))$ of time complexity, where $N$ is number of BCD digits and $b$ is the number of bits in a digit.}   \hfill $\blacksquare$ 
\\
\textbf{Proof}
The proposed BCD addition algorithm  being parallel, can be represented as a tree structure where addends are the root node of the tree $u$ and child nodes $v$ are direct logic implementation circuits, addition with 3-correction logic circuits as well as the output selection circuits.

So, a directed graph $G = (V,E)$  can be constructed where,
\begin{center}
$V \in \{u,v_1,v_2,..v_n\}$ and\\
$E \in \{(u,v_1),(u,v_2),(v1,v_2),...,(v_2,v_n)\}$.\\
\end{center}
It is obvious that, there exists exactly one pair of vertices $(u,v)$ of path length $d(u,v)$, which is the highest path length among any pair of vertices in the graph. So, the diameter of the graph $G$ is unique. Now, it is sufficient to prove that, the length of the diameter is $(log_2 b)$ where $b$ is the number of bits in a BCD digit.  

Take any node $w$ and find the vertex which is furthest from it. Now, it will be shown that, the vertex found will be either $u$ or $v$. Suppose, that the vertex found is $z$(neither $v$ nor $u$). Two cases can be considered here
\begin{enumerate}
\item suppose that $w$ is a node on path $(u,v)$. Without loss of generality, let the $(w,u)$ path have no edges overlapping with the $(w,z)$ path. So, we find the distance of the paths as follows
\begin{center}
 $d(w,z)\geq d(w,v)$.
\end{center}
But, from the shortest path algorithm (Dijkstra), we know that,
\begin{center}
$d(w,z)\geq d(w,u)$. $d(u,z)=d(u,w)+d(w,z)\geq d(u,w)+d(w,v)=d(u,v)$
\end{center}
This contradicts the assumption that, $d(u,v)$ is the unique diameter of the tree.\\
\item let $w$ does not lie on the path from $u$ to $v$. Now, either the $(w,z)$ path overlaps with the $(u,v)$ path or is disjoint. If there is overlap, consider the vertex $y$ which is the vertex closest to $w$ among the vertices which are the parts of the overlap. Without loss of generality, let the $(y,u)$ path have no edges overlapping with the $(y,z)$ path. So,
\begin{center}
$d\{y,z\}\geq d\{y,v\}$
\end{center}
From Dijkstra algorithm as we know,
\begin{center}
$d\{u,z\}=d\{u,y\}+d\{y,z\}\geq d\{u,y\}+d\{y,v\}=d\{u,v\}$
\end{center}
This once again contradicts the assumption of $(u,v)$ being the unique diameter of the tree.
\end{enumerate}
If the paths do not overlap, there are vertices $x$ and $y$ on the $(u,v)$ and $(w,z)$ paths, respectively which are closest to each other.So,
\begin{center}
 $d(y,z)\geq d(y,v)$
\end{center}
But according to Dijkstra algorithm,
\begin{center}
$d(u,z)=d(u,y)+d(y,z)\geq d(u,y)+d(y,v)>d(u,x)+d(x,v)=d(u,v)$
\end{center}
Hence, the assumption that $d\{u,v\}$ is the diameter is contradicted. In each case, we have seen that there is a contradiction if $z$ is not one of $u$ or $v$. Hence it follows that $z$, the furthest vertex from $w$, is either $u$ or $v$. So,it is proved that the furthest vertex from $u$ is $v$. Hence, while calculating the distance using DFS algorithm, we  actually find the diameter of the tree in the second run of DFS.
Since the diameter is unique, the cost of traversing from $v_1$ to $v_2$ is $log_2b$. For a $N$-digit BCD adder, the time complexity becomes $O(N(log_2 b) + (N - 1))$. \hfill $\square$

\subsection{Proposed Parallel BCD Adder Circuit Using LUT}
\begin{figure*}
\centerline{\includegraphics[scale=.6]{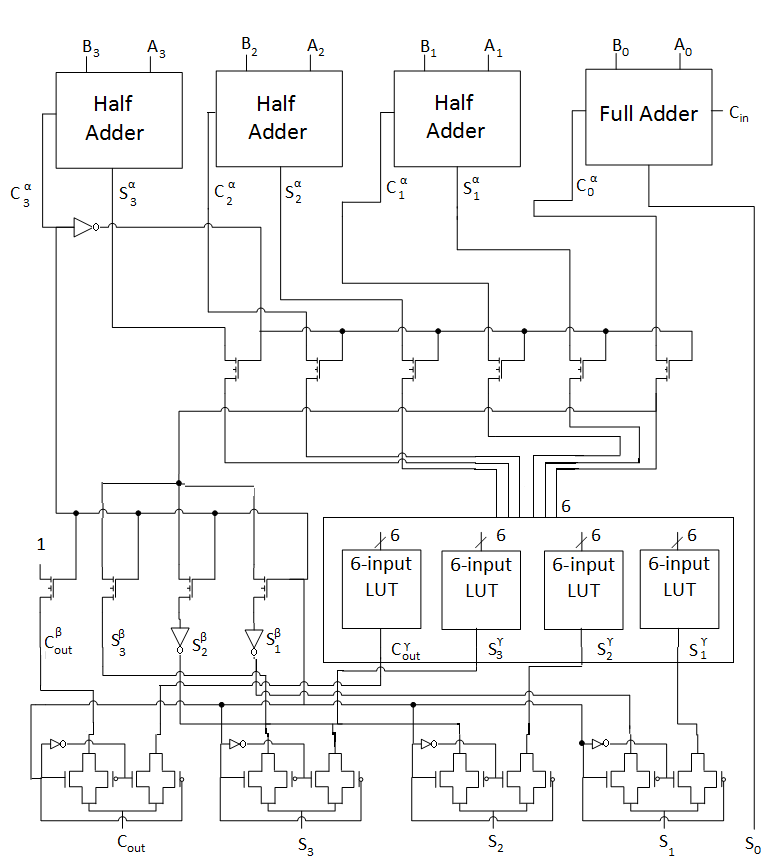}}
\caption{Proposed 1-Digit BCD Adder Circuit.}
\label{fig:add}
\end{figure*}

A LUT-based BCD adder is designed by using the proposed BCD addition algorithm and LUT architecture. An algorithm for the construction of proposed BCD adder circuit is presented in Algorithm \ref{alg:four}. According to the algorithm, the circuit is depicted in Fig. \ref{fig:add}. For the addition of the least significant bit with carry from the previous digit addition, a full adder is used. Three half-adders are used for individual bit-wise addition operation of the most significant three bits. Depending on the value of $C_3$, Equation \ref{eq: three} and Equation \ref{equ: four} are followed in the proposed circuit architecture by using the transistors and LUTs, where four number of 6-input LUTs are used to add the output from the half-adders and full adder \{$S^{\alpha}_3,...,S^{\alpha}_1, C_0$\} with the correction by adding 3, if the sum is greater than or equals to five. Depending on the value of $C_3$, a switching circuit is used to follow Equation \ref{equ: 14}. The proposed circuit gains huge delay reduction due to its parallel working mechanism compared to existing BCD adder circuits.
 
\begin{figure*}
\centerline{\includegraphics[scale = 0.45]{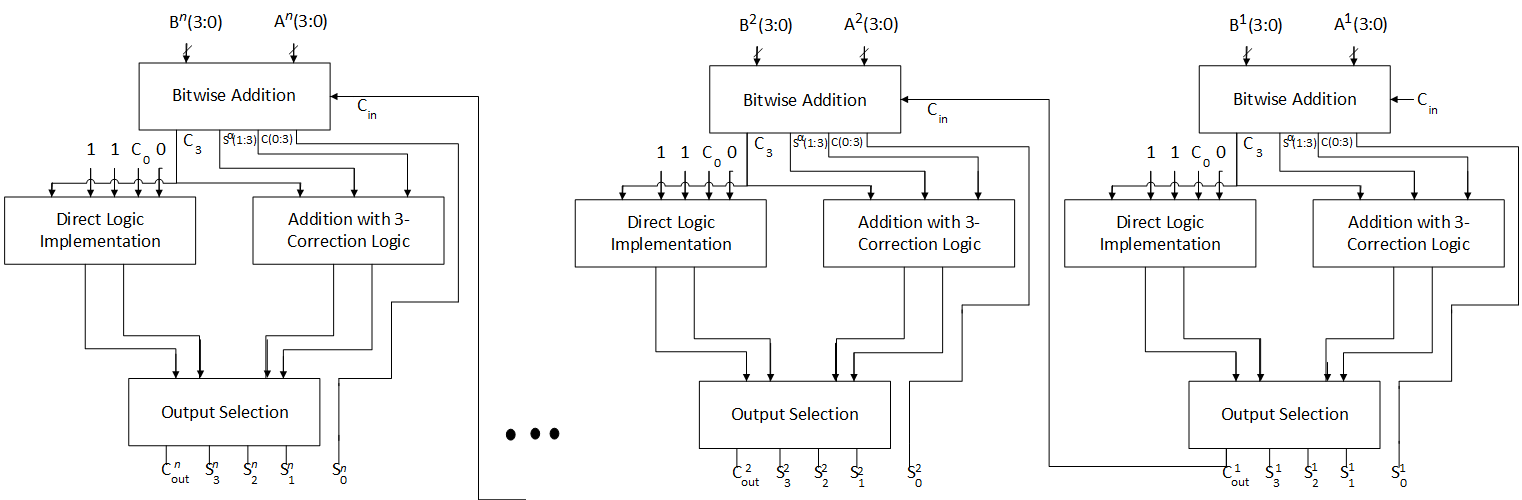}}
\caption{Block Diagram of the Proposed $N$-Digit BCD Adder Circuit.}
\label{fig:bladd}
\end{figure*}

By using the proposed 1-digit BCD adder circuit, we can easily create an $N$-digit BCD adder circuit, where the $C_{out}$  of one digit adder circuit is sent to the next digit of the BCD adder circuit as a $C_{in}$. Therefore, the generalized $N$-digit BCD adder computes sequentially by using the previous carry, the block diagram of which is shown in Fig. \ref{fig:bladd}.   
\begin{equation}
      C_{out} S_3 S_2 S_1=
      \begin{cases}
      C^{\beta}_{out} S^{\beta}_3 S^{\beta}_2 S^{\beta}_1, & \text{if}\ C_3=1 \\
      C^{\gamma}_{out} S^{\gamma}_3 S^{\gamma}_2 S^{\gamma}_1, & \text{otherwise}
      \end{cases}
      \label{equ: 14}
\end{equation}

\begin{algorithm}
\SetAlgoNoLine
\KwIn{Two 1-digit BCD numbers $A = \{A_3 A_2 A_1 A_0\}$ and $B = \{B_3 B_2 B_1 B_0\}$\;}
\KwOut{Sum $S = \{S_3 S_2 S_1 S_0\}$ and Carry = $C_{out}$\;}
     
     Apply a full adder circuit where Input:= $\{ A_0, B_0, C_{in}\}$ and Output:= $\{C_0, S_0 \}$;\\
	$i \leftarrow1$;\\
	\Repeat{ $(i=3)$}{
	Apply a half adder circuit where Input:= $\{ A_i, B_i\}$ and Output:= $\{C_i, S_i \}$;\\	
	}    
     
	 \eIf {($C_3=1$)}{\text{ $S^{\beta}_3\leftarrow C_0$; $C^{\beta}_{out}\leftarrow1$;}\\
	 \eIf{$C_0=0$}{\text{$S^{\beta}_1 \leftarrow1$; $S^{\beta}_2 \leftarrow1$;}}{\text{$S^{\beta}_1 \leftarrow0$; $S^{\beta}_2\leftarrow0$;}}}
	 { 
	 Apply four 6-input LUTs where each LUT's Input:= $\{S^{\alpha}_3, S^{\alpha}_2, S^{\alpha}_1, C_3, C_2, C_1\}$\\ and combined Output:= $\{C^{\gamma}_{out}$ $S^{\gamma}_3$ $S^{\gamma}_2$ $S^{\gamma}_1\}$;\\
	 
}
	$j \leftarrow1$;\\
	\Repeat{ $(j=3)$}{
	Apply a switching circuit where Input:= $\{ S^{\gamma}_j, S^{\beta}_j\}$ and Output:= $\{S_j \}$;\\	
	}
	
    Apply fourth switching circuit where Input:= $\{ C^{\gamma}_{out}, C^{\beta}_{out}\}$ and Output:= $\{C_{out} \}$;\\

\caption{Proposed Algorithm for the Construction of an 1-Digit BCD Adder Circuit}
\label{alg:four}
\end{algorithm}
 
\section{Simulation Results and Performance Analysis}
\label{sec: simulation}
 
As the BCD adder circuits being compared contain different types of logic gates and logic modules, it is better to preserve the basic modules as described in the architectures as long as they correspond to the commonly available cells in a typical standard cell library. The area and delay of the proposed BCD adder circuits are derived and expressed in terms of the area and critical path delay of the basic logic modules that can be found in a typical standard cell library for different operator sizes. These theoretical estimates are then calibrated by the basic logic modules from \emph{CMOS 45 nm open cell library} \cite{nangate}. Table \ref{tab: here} shows the area and critical path delay of basic logic gates. In this table, we have taken the core logic gates such as inverter, 2-input AND, OR and EX-OR gates. Table \ref{tab: there} calculates the area and critical path delay of some logic modules such as full adder, half adder and multiplexer by using the Table \ref{tab: here}. It is required to mention that, the area has been calculated in terms of number of transistors. 


\begin{table}
\centering
\caption{Area and Critical Path Delay of Basic Logic Gates}
\label{tab: here}
\scalebox{1.0}{
\begin{tabular}{|c|c|c|}
\hline
Basic Logic Gates & Area (in transistors)  & Critical Path Delay (ns) \\ \hline
Inverter (INV)         & 1                           & 1          \\ \hline
2-input AND       & 6                        & 4.68       \\ \hline
2-input OR        & 6                       & 4.5        \\ \hline
2-input EX-OR     & 8                        & 4.72       \\ \hline
\end{tabular}}
\end{table}

\begin{table}
\centering
\caption{Area and Critical Path Delay of Basic Logic Modules}
\label{tab: there}
\scalebox{1.0}{
\begin{tabular}{|c|c|c|}
\hline
Elements           & Area (in transistors)  & Critical Path Delay (ns) \\ \hline
2-to-1 Multiplexer (MUX)& 20                        & 10.18      \\ \hline
Half Adder (HA)        & 14                       & 4.72       \\ \hline
Full Adder (FA)        & 34                        & 13.9       \\ \hline

\end{tabular}}
\end{table}
 
The area complexity of the proposed BCD adder is derived from its basic logic modules. The proposed BCD adder requires three half adders, one full adder, four 6-input LUTs, six inverters and twenty six transistors. Thus, the total area of the proposed BCD adder ($A_{proposed}$) can be determined as follows: 

\begin{multline}
\label{equ: Proposedareapre}
  A_{proposed} = (3 \times A_{HA}) + (1 \times A_{FA}) + (4 \times A_{6-LUT}) + \\ (6 \times A_{INV}) + (26 \times A_{transistor})
\end{multline}

Table \ref{tab: na} shows the comparison among the proposed and existing BCD adders in terms of area. It is evident from Table \ref{tab: na} that the proposed design requires 108 transistors and four 6-input LUTs whereas the best known existing methods [10] [11] require 132 transistors and four 6-input LUTs. Thus the proposed BCD adder gains an improvement of 18.18\% in terms of area for pre-layout simulation result.  Similarly, the critical path delay of the proposed BCD adder contains one full adder, one 6-input LUT, two inverters and two transistors. Therefore, the critical path delay of the proposed BCD adder ($D_{proposed}$) can be calculated as follows:
 
\begin{multline}
\label{equ: Proposeddelaypre}
D_{proposed} = 1 \times D_{FA} + 1 \times D_{6-LUT} +   2 \times D_{INV} + \\ 2 \times D_{transistor} 
\end{multline}

Table \ref{tab: nd} shows the comparison among the proposed and existing BCD adders in terms of critical path delay. It is shown from Table \ref{tab: nd} that the proposed BCD adder requires 41.8 ns of delay whereas the best known existing methods [10] [11] require 69.56 ns of delay. Therefore the proposed BCD adder achieves an improvement of 39.9\% in terms of critical path delay in pre-layout simulation result.

\begin{table*}
\centering
\caption{Comparison of Area among the Existing and the Proposed $N$-Digit BCD Adders for Pre-Layout Simulation}
\label{tab: na}
\begin{tabular}{|c|c|c|c|}
\hline
Method                     & Area Expression                                                                                                                                                                    & Area* & LUT Count      \\ \hline
Gao et al {[}10{]} {[}11{]} & \begin{tabular}[c]{@{}c@{}}$N$ $\times$ (1 $\times$ $A_{FA}$ + 3 $\times$ $A_{MUX}$ + \\3 $\times$ $A_{Ex-OR}$  +2 $\times$ $A_{INV}$ \\+ 2 $\times$ $A_{AND}$+4 $\times$ $A_{6-LUT}$)\end{tabular} &                      132$N$  &4$N$\\ \hline
Bioul et al{[}12{]}          & $N$ $\times$ (8 $\times$ $A_{6-LUT}$ + 6 $\times$ $A_{MUX}$)                                                                                                                                    & 120$N$         &   8$N$          \\ \hline
Vazquez et al {[}13{]}          & \begin{tabular}[c]{@{}c@{}}$N$ $\times$( 5 $\times$ $A_{6-LUT}$ + 4 $\times$ $A_{MUX}$\\ + 4 $\times$ $A_{Ex-OR}$ + 2 $\times$ $A_{INV}$ +\\ 2 $\times$ $A_{AND}$)\end{tabular}         & 134$N$           &5$N$                 \\ \hline
Vazquez et al {[}14{]} &\begin{tabular}[c]{@{}c@{}}$N$ $\times$ ( 8 $\times$ $A_{6-LUT}$ + \\7 $\times$ $A_{MUX}$ + 8 $\times$ $A_{Ex-OR}$                                                                                                          ) \end{tabular} & 204$N$   & 8$N$                  \\ \hline
Proposed                   &  \begin{tabular}[c]{@{}c@{}}$N$ $\times$ (3 $\times$ $A_{HA}$ + 1 $\times$ $A_{FA}$ + \\4 $\times$ $A_{6-LUT}$ +   6 $\times$ $A_{INV}$ + \\26 $\times$ $A_{transistor}$)\end{tabular}     & 108$N$         &  4$N$ \\ \hline
\end{tabular}
\begin{tablenotes}
    \item  \hspace{30 pt} `*' Represents ``Area has been calculated in terms of transistors.''
    \end{tablenotes}

\end{table*}

\begin{table*}
\centering
\caption{Comparison of Delay among the Existing and the Proposed $N$-Digit BCD Adders for Pre-Layout Simulation}
\label{tab: nd}
\begin{tabular}{|c|c|c|}
\hline
Method                     & Delay Expression                                                                                                                                                                      & Critical Path Delay (ns)\\ \hline
Gao et al {[}10{]} {[}11{]} & \begin{tabular}[c]{@{}c@{}}$N$ $\times$ (1 $\times$ $D_{FA}$ + 2 $\times$ $D_{MUX}$ +\\  1 $\times$ $D_{Ex-OR}$ +1 $\times$ $D_{INV}$ +\\  1 $\times$ $D_{AND}$+1 $\times$ $D_{6-LUT}$)\end{tabular} & 69.56$N$               \\ \hline
Bioul et al {[}12{]}          & $N$ $\times$ (4 $\times$ $D_{6-LUT}$ + 4 $\times$ $D_{MUX}$)                                                                                                                                         & 140.72$N$              \\ \hline
Vazquez et al {[}13{]}          & \begin{tabular}[c]{@{}c@{}}$N$ $\times$ (1 $\times$ $D_{6-LUT}$ + 4 $\times$ $D_{MUX}$ + \\ 2 $\times$ $D_{Ex-OR}$ +1 $\times$ $D_{INV}$ +\\  1 $\times$ $D_{AND}$)\end{tabular}                     & 80.74$N$              \\ \hline
Vazquez et al {[}14{]}          & \begin{tabular}[c]{@{}c@{}}$N$ $\times$ (4 $\times$ $D_{6-LUT}$ + 4 $\times$ $D_{MUX}$ \\ + 6 $\times$ $D_{Ex-OR}$)\end{tabular}                                                                     & 168.64$N$              \\ \hline
Proposed                   & \begin{tabular}[c]{@{}c@{}}$N$ $\times$ (1 $\times$ $D_{FA}$ + 1 $\times$ $D_{6-LUT}$ + \\  2 $\times$ $D_{INV}$ + 2 $\times$ $D_{transistor}$)\end{tabular}                                         & 41.8$N$                \\ \hline
\end{tabular}
\end{table*}

%
%

 
\subsection{FPGA Implementation and Post-Layout Simulation Results} 

\begin{figure*}
\centerline{\includegraphics{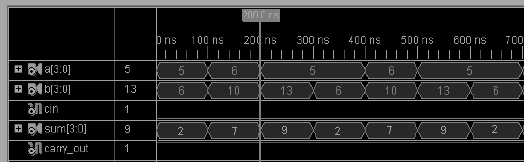}}
\caption{Simulation Result of BCD Adder with Intermediate Carry C1= 1.}
\label{fig:sc1}
\end{figure*}

\begin{figure*}
\centerline{\includegraphics{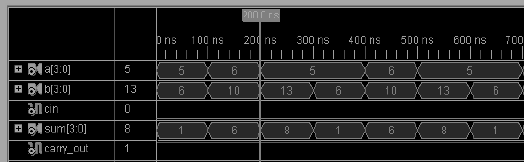}}
\caption{Simulation Result of BCD Adder with Intermediate Carry C1= 0.}
\label{fig:sc0}
\end{figure*}

The proposed BCD adder was coded in VHDL and
implemented in a Virtex-6 XC6VLX75T Xilinx FPGA with a -3 speed grade using by ISE 13.1. The results are compared with the earlier approaches proposed in
[11]-[14] by using the same experimental setup for fair comparison. The delays were extracted from
Postplacement-and-Routing Static Timing Report and the
LUTs usage was obtained from Place-and-Routing Report. Besides, the simulations of the proposed BCD adder are demonstrated in Fig. \ref{fig:sc1} and Fig. \ref{fig:sc0} with carry 1 and 0, respectively.

\begin{figure}
\centerline{\includegraphics[ width = 9 cm]{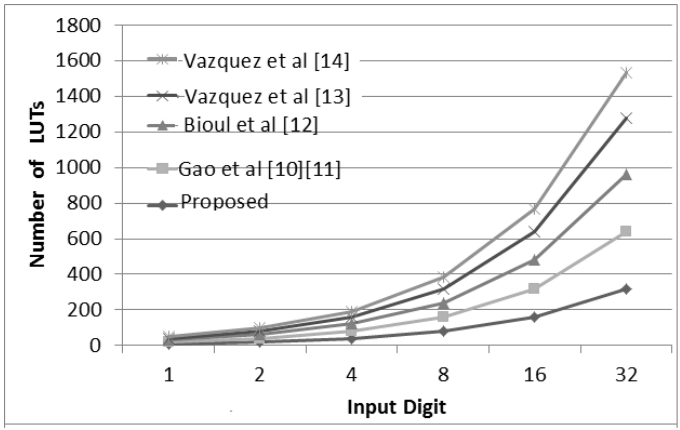}}
\caption{Graphical Analysis of Area of Existing and Proposed BCD Adder Circuits for Post-Layout Simulation.}
\label{fig:comaa}
\end{figure}

\begin{figure}
\centerline{\includegraphics[ width = 9 cm]{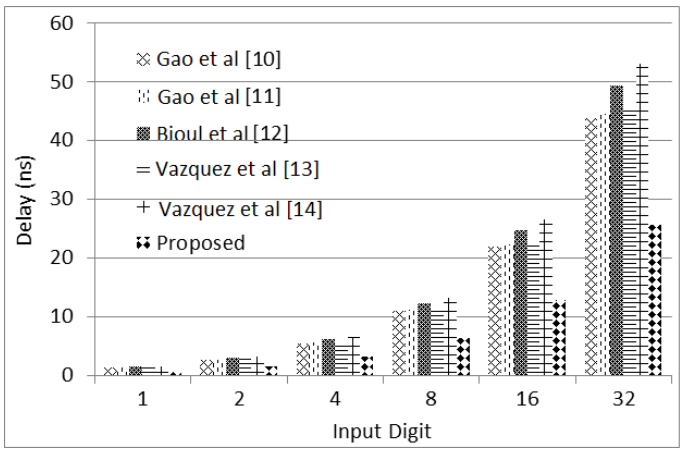}}
\caption{Graphical Analysis of Delay of Existing and Proposed BCD Adder Circuits for Post-Layout Simulation.}
\label{fig:comad}
\end{figure}

\begin{figure}
\centerline{\includegraphics[width = 9 cm]{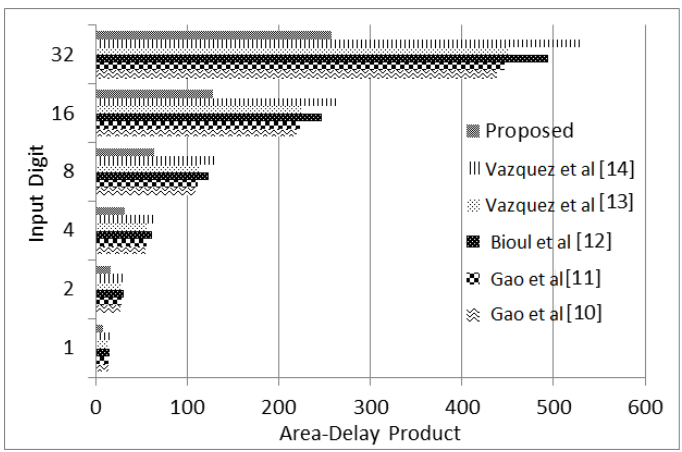}}
\caption{Graphical Analysis of Area-Delay Product of Existing and Proposed BCD Adder Circuits.}
\label{fig:comap}
\end{figure}

The proposed BCD adder is high-speed due to its less time complexity with optimum critical path delay and cost-efficient due to its area and area-delay product efficiency. Comparison of area, delay and area-delay product among existing [11]-[14] and the proposed BCD adder circuits for various number of input digits are shown in graphical representation in Fig. \ref{fig:comaa}, Fig. \ref{fig:comad} and Fig. \ref{fig:comap}, respectively with improvement of 20\%, 41.32\% and 53.06\% in terms of area, delay and area-delay product, respectively compared to the existing best method [10] [11]. It is to be noted that, the results shown in Fig. \ref{fig:comaa}, Fig. \ref{fig:comad} and Fig. \ref{fig:comap} for earlier approaches [11]-[14] have been re-implemented by using Virtex-6 platform.

\section{Conclusion}
\label{sec: concl}
In twenty years, reconfigurable computing has grown from a wild, exploratory idea to a viable alternative to Application-Specific Integrated Circuits (ASICs) and fixed microprocessors in our computing systems. Besides, BCD (Binary Coded Decimal) addition being the basic arithmetical operation, it is the main focus. The proposed BCD adder is highly parallel, which mitigates the significant carry propagation delay of addition operation. The proposed BCD adder circuit is not only faster but also area-efficient compared to the existing best known circuit. The pre-layout simulation provides 18.18\% and 39.9\% efficiency in terms of area and critical path delay reduction, respectively compared to the existing best known BCD adder circuit. The proposed BCD adder circuit is simulated using Xilinx Virtex-6. The correctness and efficiency of the circuit is proved in the proposed section and simulation section using corresponding tables, figures and lemma. It is shown by the comparative analysis that the proposed BCD adder is 20\% and 41.3\% improved in terms of area and delay, respectively compared to the existing best known adder circuit along with 53.06\% improvement in area-delay product. These improvements in FPGA-based BCD addition will consequently influence the advancement in computation and manipulation of decimal digits, as it is more convenient to convert from decimal to BCD than binary. Besides, FPGA implementation will be beneficial to be applied in bit-wise manipulation, private key encryption and decryption acceleration, heavily pipe-lined and parallel computation of NP-hard problems, automatic target generation and many more applications \cite{24} \cite{21}.


\section*{Acknowledgment}
Zarrin Tasnim Sworna and Mubin Ul Haque has been granted fellowship from the Ministry of Information and Technology, People's Republic of Bangladesh under the program of higher studies and research with the reference no. 56.00.0000.028.33.058.15-629.



%

\end{document}